%% file: main.tex
\pdfoutput=1
\documentclass{article}
\usepackage{spconf,amsmath,amssymb,graphicx}
\usepackage{color}
\usepackage{bm}
\usepackage{epstopdf}

\usepackage[farskip=0pt]{subfig}
\usepackage{algorithm}
\usepackage{algorithmic}
\usepackage{cite}

\def\n{{\mathbf n}}

\def\x{{\mathbf x}}
\def\y{{\mathbf y}}
\def\l{{\mathbf l}}

\def\hx{{\hat{\mathbf x}}}

\def\A{{\mathbf A}}
\def\B{{\mathbf B}}
\def\C{{\mathbf C}}
\def\D{\mathbf{D}}

\def\H{{\mathbf H}}

\def\L{{\mathbf L}}

\def\Q{{\mathcal Q}} 

\def\W{{\mathbf W}}

\def\bS{{\mathbf S}}

\def\cG{{\cal G}}
\def\cV{{\cal V}}
\def\cE{{\cal E}}
\def\cN{{\cal N}}

\def\sR{{\mathbb R}}

\def\hT{{\hat T}}

\def\ie{\emph{i.e.}}

\DeclareMathOperator*{\diag}{diag} 
\DeclareMathOperator*{\zeros}{zeros} 
\DeclareMathOperator*{\argmin}{argmin} 

\title{Reconstruction-Cognizant Graph Sampling using Gershgorin Disc Alignment}
\name{Yuanchao Bai$^{\star}$, Gene Cheung$^{\dagger}$, Fen Wang$^{\ddagger}$, Xianming Liu$^{\$}$, Wen Gao$^{\star}$}
\address{
$^{\star}$Peking University, China\;\;\;
$^{\dagger}$York University, Canada\\
$^{\ddagger}$Xidian University, China\;\;\;
$^{\$}$Harbin Institute of Technology, China
}
\graphicspath{{figures/}}

\begin{document}
\ninept
\maketitle
\begin{abstract}
Graph sampling with noise is a fundamental problem in graph signal processing (GSP).
Previous works assume an unbiased least square (LS) signal reconstruction scheme and select samples greedily via expensive extreme eigenvector computation.
A popular biased scheme using graph Laplacian regularization (GLR) solves a system of linear equations for its reconstruction.
Assuming this GLR-based scheme, we propose a reconstruction-cognizant sampling strategy to maximize the numerical stability of the linear system---\textit{i.e.}, minimize the condition number of the coefficient matrix.
Specifically, we maximize the eigenvalue lower bounds of the matrix, represented by left-ends of Gershgorin discs of the coefficient matrix.
To accomplish this efficiently, we propose an iterative algorithm to traverse the graph nodes via Breadth First Search (BFS) and align the left-ends of all corresponding Gershgorin discs at lower-bound threshold $T$ using two basic operations: disc shifting and scaling.
We then perform binary search to maximize $T$ given a sample budget $K$.
Experiments on real graph data show that the proposed algorithm can effectively promote large eigenvalue lower bounds, and the reconstruction MSE is the same or smaller than existing sampling methods for different budget $K$ at much lower complexity.
\end{abstract}
%

\section{Introduction}
\label{sec:intro}
\input{intro}

\section{Preliminaries}
\label{sec:reconstruction}
\input{reconstruction}

\vspace{-0.05in}
\section{Reconstruction-Cognizant Sampling}
\label{sec:sampling}
\input{sampling}

\section{Experiments}
\label{sec:experiments}
\input{experiments}

\section{Conclusion}
To address the ``graph sampling with noise" problem, in this paper we propose a reconstruction-cognizant graph sampling scheme that assumes a biased reconstruction based on graph Laplacian regularization (GLR) and maximizes the stability of the solution's linear system.
In particular, our proposed BS-BFIS promotes large lower-bounds of $\lambda_{\min}$ via Gershgorin disc alignment.
Besides stability of signal reconstruction, the proposed algorithm leads to same or better reconstruction MSE against existing methods at lower complexity. 






\vfill\pagebreak

\bibliographystyle{IEEEbib}
\bibliography{ref2,graph_refs}

\end{document}

%% file: intro.tex
Graph sampling is a basic problem in \textit{Graph Signal Processing} (GSP) \cite{shuman13, ortega2018ieee, gene2018ieee}.
While the noiseless sampling case is extensively studied \cite{sampling2008TAMS,sampling2014icassp,e_optimal2015,MFN2016TSP,sp_proxy2016,2018greedy,r_sampling2018ACHA}, the ``sampling with noise'' case remains challenging.
Previous works typically assume an unbiased \textit{least square} (LS) signal reconstruction scheme from sparse samples \cite{MFN2016TSP,2018greedy,MIA2018Fen}, which leads to a minimum mean square error (MMSE) formulation and the known A-optimality criterion for independent additive noise \cite{convex_optimization}.
The criterion is minimized greedily per sample using schemes that compute extreme eigenvectors \cite{MFN2016TSP,2018greedy}, which is not scalable for large graphs.
(\cite{MIA2018Fen} does not compute eigenvectors, but computes many matrix-vector multiplications for good approximation.)


Instead of unbiased LS reconstruction, recent biased graph signal restoration schemes employ signal priors, including \textit{graph Laplacian regularization} (GLR) \cite{GDenoising2017Pang,SoftDecode2017XLiu} and \textit{graph total variation} (GTV) \cite{DTV2013SIAM,GTV2017,deblur2017YBai}.
In particular, biased schemes using GLR solve a system of linear equations for signal reconstruction via fast numerical methods like \textit{conjugate gradient} (CG) \cite{hestenes1952methods}.
In this paper, assuming a GLR signal reconstruction scheme, we propose a \textit{reconstruction-cognizant} sampling strategy to maximize the numerical stability of the linear system---\textit{i.e.}, minimize the condition number (ratio of the largest to smallest eigenvalues) of the coefficient matrix.
By coupling the GLR reconstruction method to sampling during optimization, we expect a better-performing sample set that yields higher quality when the sampling and reconstruction schemes are deployed in tandem.

Computing the extreme eigenvalues of a large matrix directly is expensive, using prevalent methods such as implicitly restarted Arnoldi method \cite{ARPACK} or the Krylov-Schur algorithm \cite{eigenvalue2002SIAM}.
Instead, we maximize the minimum of all eigenvalue lower bounds of the matrix, where each bound is represented by the left-end of a Gershgorin disc of the coefficient matrix \cite{linear_algebra}.
We introduce two basic operations to manipulate a Gershgorin disc: disc shifting via sampling, and disc scaling via similarity transform.
We design a \textit{Breadth First Iterative Sampling} (BFIS) algorithm to traverse all nodes via \textit{Breath First Search} (BFS), and align the left-ends of all discs to a lower bound threshold $T$.
We then perform binary search (BS) to maximize $T$ given a sampling budget $K$.
\textit{Note that unlike existing greedy sampling schemes \cite{e_optimal2015,MFN2016TSP,sp_proxy2016,2018greedy}, our scheme never explicitly computes extreme eigenvectors, and thus can scale gracefully to very large graphs.}
Experiments on both illustrative examples and real graph data
demonstrate that our proposed BS-BFIS algorithm promotes large eigenvalue lower bounds, and the reconstruction MSE is the same or smaller than existing sampling methods \cite{e_optimal2015,sp_proxy2016,MIA2018Fen} for different budget $K$.


%% file: reconstruction.tex
We define a graph $\cG$ as a triplet $\cG(\cV,\cE,\W)$, where $\cV$ and $\cE$ represent sets of $N$ nodes and $M$ edges in the graph, respectively.
Associated with each edge $(i, j)\in\cE$ is a weight $w_{i,j}$, which reflects the correlation or similarity between two nodes $i$ and $j$.
We assume a connected undirected graph;  \textit{i.e.}, $w_{i,j}=w_{j,i}, \forall i, j \in \cV$.
$\W$ is an \textit{adjacency} matrix with $w_{i,j}$ as the $(i,j)$-th entry of the matrix.
Typically, $w_{i,j}>0$ for $\forall (i, j)\in\cE$, and $w_{i,j}=0$ otherwise.

Given $\W$, the \textit{combinatorial graph Laplacian} matrix $\L$ is computed as \cite{ortega2018ieee}:
\begin{equation}
    \L\triangleq \D-\W
    \label{eq:graph_laplacian}
\end{equation}
where $\D=\diag(\W\mathbf{1})$ is a diagonal \textit{degree} matrix. $\mathbf{1}$ is a vector of all 1's and $\diag(\cdot)$ is an operator that returns a diagonal matrix with the elements of an input vector on the main diagonal.


%
\textit{Graph Laplacian regularizer} (GLR) \cite{GDenoising2017Pang} is a smoothness prior for signals on graphs, which has demonstrated its effectiveness in numerous applications, such as semi-supervised learning \cite{gl_semi_learning2004lt,su2018}, image processing \cite{GDenoising2017Pang,SoftDecode2017XLiu,gene2018ieee} and computer graphics \cite{Laplacian_mesh2005eurographics}.
Given observation $\y$ on a graph $\cG$, one can formulate an optimization for the target signal $\hat{\x} \in \mathbb{R}^N$ using GLR as follows:
\begin{equation}
    \hx=\argmin_\x \|\H\x-\y\|^2_2 +
    \mu \; \x^{\top} \L \, \x
    \label{eq:reconstruction}
\end{equation}
where $\H$ represents a signal degradation process.
$\mu$ is a tradeoff parameter to balance GLR against the $l_2$-norm data fidelity term.

In this work, we focus on signal reconstruction from sparse samples.
The observation model for signal samples can be modeled linearly as follows \cite{sampling2008TAMS,sampling2014icassp,e_optimal2015,MFN2016TSP,sp_proxy2016,r_sampling2018ACHA,2018greedy,MIA2018Fen}:
\begin{equation}
    \y=\H\x+\n
    \label{eq:sampling}
\end{equation}
where $\H\in\sR^{K\times N}$ is a sampling matrix \cite{MIA2018Fen}.
$\x\in\sR^N$ is an original graph signal, and $\y\in\sR^K$, $0<K<N$, is a sampled signal of dimension $K$ corrupted by additive noise $\n$.

Since objective (\ref{eq:reconstruction}) is quadratic, the optimal solution can be obtained by solving a system of linear equations:
\begin{equation}
    (\H^{\top}\H+\mu\L)\x=\H^{\top} \y.
    \label{eq:linear_equation}
\end{equation}
Because both $\H^{\top}\H$ and $\L$ are singular matrices,
(\ref{eq:linear_equation}) can potentially be poorly conditioned.
From this observation, we next study the impact of sampling on the numerical stability of (\ref{eq:linear_equation}) and propose a reconstruction-cognizant sampling strategy.

%% file: sampling.tex
\subsection{Graph Sampling and Reconstruction Stability}
%
Reconstructing a sampled signal with GLR leads to solving a linear equation (\ref{eq:linear_equation}).
Denote by a diagonal matrix $\A=\H^{\top}\H\in\sR^{N\times N}$ satisfying
\begin{equation}
    a_{i,i}=\begin{cases}
    1,~~~~~i\in \Phi,\\
    0,~~~~~\mbox{otherwise}.
    \label{eq:ata}
    \end{cases}
\end{equation}
where $\Phi$ is a set of indices of sampled nodes.
Denote by $\B=\A+\mu\L$.
From \textit{Gershgorin Circle Theorem} (GCT)\footnote{https://en.wikipedia.org/wiki/Gershgorin\_circle\_theorem}, each eigenvalue $\lambda$ of $\B$ lies within one \textit{Gershgorin disc} $\Psi_i(b_{i,i},R_i)$ with disc center $b_{i,i}$ and radius $R_i$, \emph{i.e.},
\begin{equation}
    b_{i,i} - R_i \le \lambda\le b_{i,i} + R_i,
    \label{eq:disk}
\end{equation}
where $R_i=\sum _{j\ne i} |b_{i,j}|=\mu\sum _j w_{i,j}=\mu d_{i}$, and $d_i$ is the degree of node $i$. The second equation is true since there are no self-loops in $\cG$.
Center of disc $i$ is $b_{i,i} = \mu d_i+a_{i,i}$.

The upper bound of all eigenvalues can be computed as:
\begin{equation}
    \max_i \{b_{i,i} + R_i\}=\max_i \{a_{i,i} + 2\mu d_i\}\le1+2\mu \max_i d_i.
\end{equation}
For a sparse graph with maximum degree $d_{\max}$ for each node, the eigenvalue upper bound is $1 + 2 \mu \, d_{\max}$, which is not large.

The lower bound of all eigenvalues is computed as:
\begin{equation}
    \min_i \{b_{i,i} - R_i\}=\min_i a_{i,i}=0.
\end{equation}
In words, for each unsampled node, its Gershgorin disc in $\B$ has left-end at $0$---an eigenvalue lower bound at $0$.
Thus the minimum eigenvalue $\lambda_{\min}$ of $\B$ can also be close to the $0$ lower bound, severely magnifying the condition number $\lambda_{\max}/\lambda_{\min}$ of $\B$, and resulting in a poorly-conditioned signal reconstruction using (\ref{eq:linear_equation}).
The extreme case is when no nodes are sampled, \emph{i.e.}, $\B=\mu\L$, and $\lambda_{\min} = 0$.
Ideally then, we would \textit{shift} all Gershgorin discs right to maximize the minimum eigenvalue lower bounds.

Via GCT, we see that we can estimate the degree of numerical instability of GLR signal reconstruction \textit{without} computing actual eigenvalues, by examining left-ends of Gershgorin discs.
We next introduce two operations to manipulate Gershgorin discs, which leads to a sampling algorithm to maximize the lower-bounds of $\lambda_{\min}$.

\subsection{Graph Sampling to Maximize Lower-bounds of $\lambda_{\min}$}
%

We first state the following linear algebra fact without proof, which we use to enable scaling of Gershgorin discs.

\vspace{0.05in}
\noindent \textbf{Fact 1}: \textit{Similarity transform} $\bS$ of a square matrix $\B$ to $\C$, defined as
\begin{equation}
    \C=\bS\B\bS^{-1},
    \label{eq:similar_trans}
\end{equation}
preserves the eigenvalues of $\B$, assuming $\bS$ is a nonsingular matrix.
\vspace{0.01in}

Using Fact 1, we will employ a diagonal $\bS$ to scale Gershgorin discs of $\B$, so that left-ends of Gershgorin discs of resulting transformed $\C$ are moved right, maximizing lower bounds of $\lambda_{\min}$.
\textit{
By scaling each disc $\Psi_i$ to move its left-end $b_{i,i} - R_i$ to the right \textit{without} affecting eigenvalues of $\B$, we are tightening one lower bound for $\lambda_{\min}$ of $\B$ per scaling operation, which helps us make more informed sampling decisions for other nodes $j\neq i$.
}

\subsubsection{Breadth First Iterative Sampling}

We introduce two basic operations to manipulate Gershgorin discs.
The first operation is \textit{disc shifting} via sampling.
As discussed, the left-end $b_{i,i} - R_i = a_{i,i}$ of the $i$-th Gershgorin disc $\Psi_i$ in matrix $\B$ shifts from 0 to 1 when node $i$ is sampled.

The second operation is \textit{disc scaling} via similarity transform.
We specify the $i$-th diagonal term $s_i$ of $\bS$ in (\ref{eq:similar_trans})---and its corresponding element $s_i^{-1}$ in $\bS^{-1}$---to scale the radius $R_i$ of $\Psi_i$ and the radii of its neighbors' discs $\Psi_j$, where $j \in \mathcal{N}_i = \{j ~|~ w_{i,j} > 0\}$.
For example, if we \textit{expand} $R_i$ using scalar $s_i>1$, then we also \textit{shrink} its neighbors' discs with $s_i^{-1}<1$.
Since $s_i$ is always offset by $s_i^{-1}$ on the main diagonal, the center $b_{i,i}$ of disc $\Psi_i$ is unchanged.


Given graph $\cG$ and an eigenvalue lower-bound threshold $T$, where $T < 1$, we apply disc shifting and scaling operations iteratively to align discs' left-ends at $T$.
The algorithm is as follows.
First, we sample a chosen node $i$ (thus moving the corresponding disc $\Psi_i$'s center $b_{i,i}$ from $\mu d_i$ to $1 + \mu d_i$, and $\Psi_i$'s left-end $a_{i,i}$ from $0$ to $1$).
Then we apply scalar $s_i$ to expand $\Psi_i$'s radius $R_i$ and align its left-end at exactly $T$.
Scalar $s_i$ must hence satisfy
\begin{equation}
    a_{i,i}+\mu \, \left(d_i-s_i\cdot\sum_{j\in \mathcal{N}_i}\frac{w_{i,j}}{s_j}\right)=T,
    \label{eq:scale_function}
\end{equation}
where initially $s_j=1$ for $j \neq i$. Solving for $s_i$ in (\ref{eq:scale_function}), we get
\begin{equation}
    s_i=\frac{a_{i,i}+\mu \, d_i-T}{\mu \, \sum_{j\in \cN_i} \frac{w_{i,j}}{s_j}}.
    \label{eq:scale_factor}
\end{equation}

Using scalar $s_i$ means we also shrink node $i$'s neighbors' discs $\Psi_j$'s radii due to $s_i^{-1}$.
Specifically, left-end $b_{j,j} - R_j$ of a neighbor $j$'s disc $\Psi_j$ ($a_{j,j}=0$) is now:
\begin{equation}
b_{j,j} - R_j = a_{j,j} + \mu \left( d_j - s_j \cdot \sum_{k \in \mathcal{N}_j \setminus \{i\}} \frac{w_{j,k}}{s_k}
- s_j \cdot \frac{w_{j,i}}{s_i} \right)
\end{equation}

If a neighboring disc $\Psi_j$'s left-end is larger than $T$, then we need not sample node $j$ and instead expand its radius to align its left-end at $T$ using (\ref{eq:scale_factor}).
This shrinks the discs of node $j$'s neighbors, and so on.
$s_j$ decreases with hops away from the sampled node.

If the left-end of $\Psi_j$ is smaller than $T$, then we sample this node ($a_{j,j}=1$) and select scalar $s_j$ using (\ref{eq:scale_factor}) again, and the process repeats.
Since we always expand a current disc ($s_i > 1$) leading to shrinking of neighboring discs (${s_i}^{-1} < 1$) in each step, the left-end of each scaled node remains larger than or equal to lower-bound $T$.
We traverse all the nodes using \textit{Breadth First Search} (BFS).
Thus, we name our proposed algorithm \textit{Breadth First Iterative Sampling} (BFIS). The BFIS is sketched in Algorithm~1.

\begin{algorithm}[!t]
\label{al:1}
\caption{Breadth First Iterative Sampling}
\begin{algorithmic}[1] 
\small
\REQUIRE  
Graph $\cG$, lower-bound $T$, the start node $i$ and $\mu$.\\
\STATE
Initialize $\D=\diag(\W\mathbf{1})$ and $\bS=\diag(\mathbf{1})$. \\
\STATE Initialize  $\A=\zeros(N,N)$, $N=|\cV|$. \\
\STATE Initialize an empty set $\Q$ for enqueued nodes. \\
\STATE Initialize an empty $queue$. \\
\STATE
$Enqueue(queue,i)$ and $\Q\leftarrow\Q\cup\{i\}$.\\
\STATE\textbf{while} $queue$ is not empty \textbf{do}\\
\STATE\ \ \ \ \ \ $k\leftarrow$Dequeue($queue$). \\
\STATE\ \ \ \ \ \ Update $s_k$ using (\ref{eq:scale_factor}).\\
\STATE\ \ \ \ \ \ \textbf{if} $s_k<1$ \textbf{do}\\
\STATE\ \ \ \ \ \ \ \ \ \ Sampling node $k$ by setting $a_{k,k}=1$.\\
\STATE\ \ \ \ \ \ \ \ \ \ Update $s_k$ using (\ref{eq:scale_factor}).\\
\STATE\ \ \ \ \ \ \textbf{endif}\\
\STATE\ \ \ \ \ \ \textbf{for} $t$ \textbf{in} $k$'s neighbours $\cN(k)$ \textbf{do}\\
\STATE\ \ \ \ \ \ \ \ \ \ \textbf{if} $t\notin\Q$ \textbf{do}\\
\STATE\ \ \ \ \ \ \ \ \ \ \ \ \ \ $Enqueue(queue,t)$ and $\Q\leftarrow\Q\cup\{t\}$.\\
\STATE\ \ \ \ \ \ \ \ \ \ \textbf{endif}\\
\STATE\ \ \ \ \ \ \textbf{endfor}\\
\STATE\textbf{endwhile}\\

\ENSURE  Sampling matrix $\A$.
\end{algorithmic}
\end{algorithm}


\vspace{-0.1in}
\subsubsection{Illustrative Example}

\begin{figure}[!t]
\centering
\includegraphics[width=0.75\linewidth]{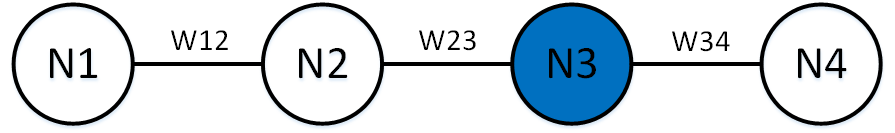}
%
%
%
\caption{An illustrative example of a 4-node line graph.}
\label{fig:nodes}
\end{figure}

\begin{figure}[!t]
\centering
\subfloat[]{
\label{fig:adj_matrix}
\includegraphics[width=0.32\linewidth]{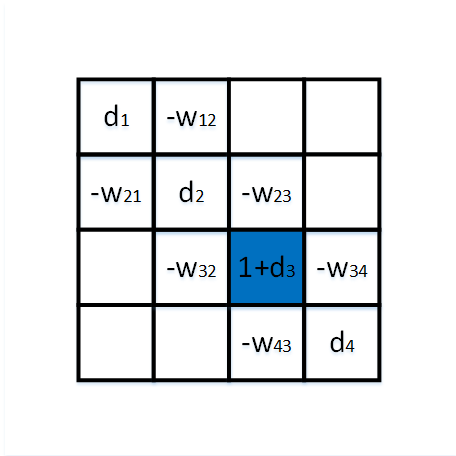}}
\subfloat[]{
\label{fig:scale}
\includegraphics[width=0.32\linewidth]{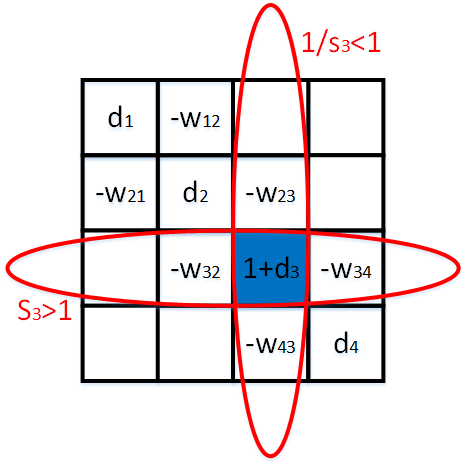}}
\subfloat[]{
\label{fig:rescale}
\includegraphics[width=0.32\linewidth]{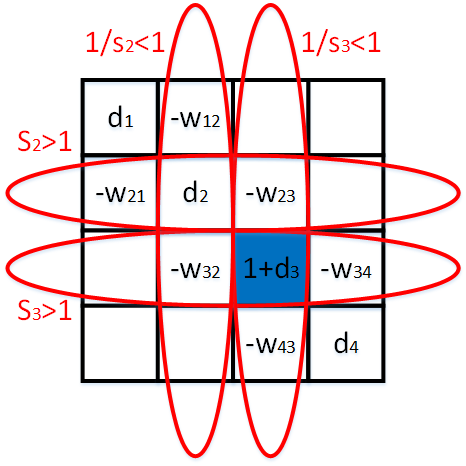}} \\
\subfloat[]{
\label{fig:disk_al}
\includegraphics[width=0.3\linewidth]{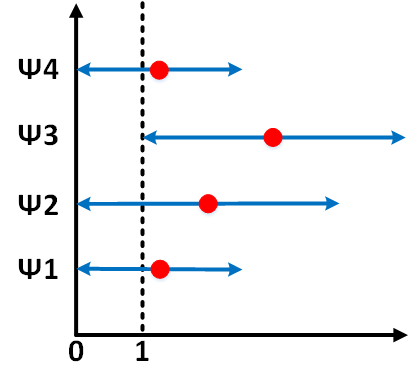}}
\subfloat[]{
\label{fig:disk_als}
\includegraphics[width=0.33\linewidth]{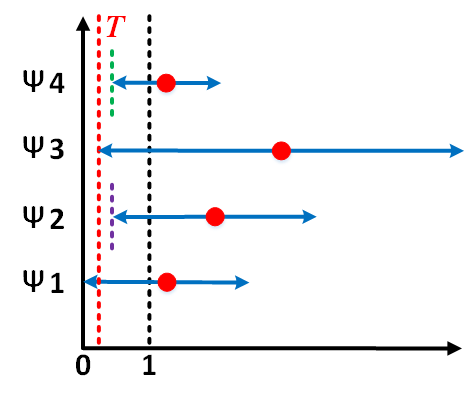}}
\subfloat[]{
\label{fig:disk_alsr}
\includegraphics[width=0.33\linewidth]{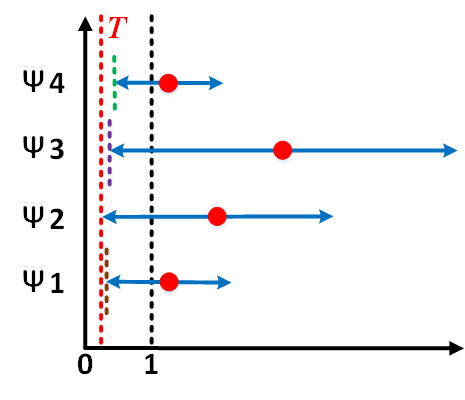}}
\caption{An illustration of BFIS. (a) sampling node $3$. (b) scaling node $3$. (c) scaling nodes $2$ and $3$. (d) Discs after sampling node $3$. (e) Discs after scaling node $3$. (f) Discs after scaling nodes $2$ and $3$.
}
\label{fig:scale_op}
\end{figure}

We use a simple example to illustrate how BFIS works. We assume a four-node graph as shown in Fig.\;\ref{fig:nodes}.
We start by sampling node $3$.
Assuming $\mu=1$, the graph's coefficient matrix $\B$ with $(3,3)$-th entry updated is shown in Fig.\;\ref{fig:adj_matrix}.
Correspondingly, left-end of node 3's Gershgorin disc---red dots and blue arrows represent disc centers and radii respectively---shifts from $0$ to $1$, as shown in Fig.~\ref{fig:disk_al}.

We next perform disc scaling on sampled node 3.
As shown in Fig.\;\ref{fig:scale}, scalar $s_3$ is applied to the third row of $\B$, and thus the radius of disc $\Psi_3$ is expanded by $s_3$ where $s_3>1$.
Simultaneously, scalar $s_3^{-1}$ is applied to the third column, and thus the radii of discs $\Psi_2$ and $\Psi_4$ are shrunk due to the scaling of $w_{2,3}$ and $w_{4,3}$ by $s_3^{-1}$.
Note that the $(3,3)$-th entry of $\B$ (and $\Psi_3$'s disc center) is unchanged, since scalar $s_3$ is offset by $s_3^{-1}$.
We see that by expanding the disc of sampled node $3$, the left-ends of discs of its neighboring nodes (nodes $2$ and $4$) shift beyond threshold $T$, as shown in Fig.\;\ref{fig:disk_als}.

We next apply scalar $s_2$ to disc $\Psi_2$ to expand its radius by $s_2$, where $s_3>s_2>1$, and the radii of discs $\Psi_1$ and $\Psi_3$ are shrunk due to the scaling of $w_{1,2}$ and $w_{3,2}$ by $s_2^{-1}$, as shown in Fig.\;\ref{fig:rescale}.
$s_2$ must be smaller than $s_3$ for the left-end of $\Psi_2$ not to move past $0$.
The discs are shown in Fig.~\ref{fig:disk_alsr}.
Subsequently, similar disc operations can be performed on $\Psi_1$ and $\Psi_4$.
Finally, the left-ends of all discs move beyond threshold $T$.

\subsubsection{Binary Search with BFIS}

Given a sample budget $K$, we perform binary search to maximize the lower-bound threshold $T$.
We call the algorithm \textit{Binary Search with BFIS} (BS-BFIS), as outlined in Algorithm~2.
At each iteration, if the number of sampled nodes in $\A$ output from BFIS is larger than $K$, then threshold $T$ is set too large, and we update $right$ to reduce $T$. On the other hand, if the number of sampled nodes is smaller than or equal to $K$, then threshold $T$ may be too small, and we update $left$ to increase $T$.
When $right-left\le\epsilon$, BS-BFIS converges and we find the maximum lower bound $\hT$ with numerical error lower than $\epsilon$.
We run BFIS again with $\hT$ to compute the $K$ sampled nodes.

\begin{algorithm}[!t]
\label{al:2}
\caption{Binary Search with BFIS}
\begin{algorithmic}[1] 
\small
\REQUIRE  
Graph $\cG$, sample size $K$, numerical precision $\epsilon$, the start node $i$ and weight parameter $\mu$.\\
\STATE Initialize $left=0$, $right=1$. \\
\STATE\textbf{while} $right-left>\epsilon$ \textbf{do}\\
\STATE\ \ \ \ \ \ $T\leftarrow(left+right)/2$. \\
\STATE\ \ \ \ \ \ $\A\leftarrow\mathbf{BFIS}(\cG,T,i,\mu)$.\\
\STATE\ \ \ \ \ \ $m\leftarrow$ the number of nodes sampled in $\A$.\\
\STATE\ \ \ \ \ \ \textbf{if} $m>K$ \textbf{do}\\
\STATE\ \ \ \ \ \ \ \ \ \ $right\leftarrow T$ \\
\STATE\ \ \ \ \ \ \textbf{else} \\
\STATE\ \ \ \ \ \ \ \ \ \ $left\leftarrow T$ \\
\STATE\ \ \ \ \ \ \textbf{endif}\\
\STATE\textbf{endwhile}\\
\STATE $\hT\leftarrow left$.
\STATE$\A\leftarrow\mathbf{BFIS}(\cG,\hT,i)$.\\
\ENSURE  Sampling matrix $\A$, maximum lower-bound $\hT$.
\end{algorithmic}
\end{algorithm}

Because the proposed BFIS executes BFS once on a graph $\cG$, the time complexity of BFIS is $O(|\cV|+|\cE|)$.
In order to achieve numerical precision $\epsilon$ in BS-BFIS, we need to employ BFIS $O(\log\frac{1}{\epsilon})$ times. Thus, the time complexity for BS-BFIS is $O\left((|\cV|+|\cE|)\log\frac{1}{\epsilon}\right)$.

%

%% file: experiments.tex
\subsection{Experimental Setting}

We apply the proposed sampling algorithm on both an illustrative line graph and a real \textit{U.S. Climate Normals database} \cite{climate_database}. 
We compare with several existing graph sampling methods: E-optimal \cite{e_optimal2015}, spectral proxies \cite{sp_proxy2016}, and MIA \cite{MIA2018Fen}.
All algorithms are implemented and run on Matlab R2015a platform.

To run BS-BFIS algorithm, there are three parameters we need to set besides graph $\cG$ and sample size $K$, \ie, numerical precision $\epsilon$, the start node $i$ and tradeoff parameter $\mu$. In experiments, we set the numerical precision $\epsilon=10^{-4}$.
Because BS-BFIS employs BFS to visit all the graph nodes, the start node $i$ determines the visiting order and affects the performance of BS-BFIS, especially when $K\ll N$.
To demonstrate the best performance of BS-BFIS, we choose the start node $i$ that leads to the largest $\hT$ via brute-force search.
For the sake of speed, the start node $i$ can be chosen randomly in practice.
In experiments, we set the tradeoff parameter $\mu$ in (\ref{eq:scale_function}) and (\ref{eq:linear_equation}) to $0.01$ for signal reconstruction.

\begin{figure}[!t]
\centering
\subfloat[]{
\label{fig:lg_bf}
\includegraphics[width=0.49\linewidth]{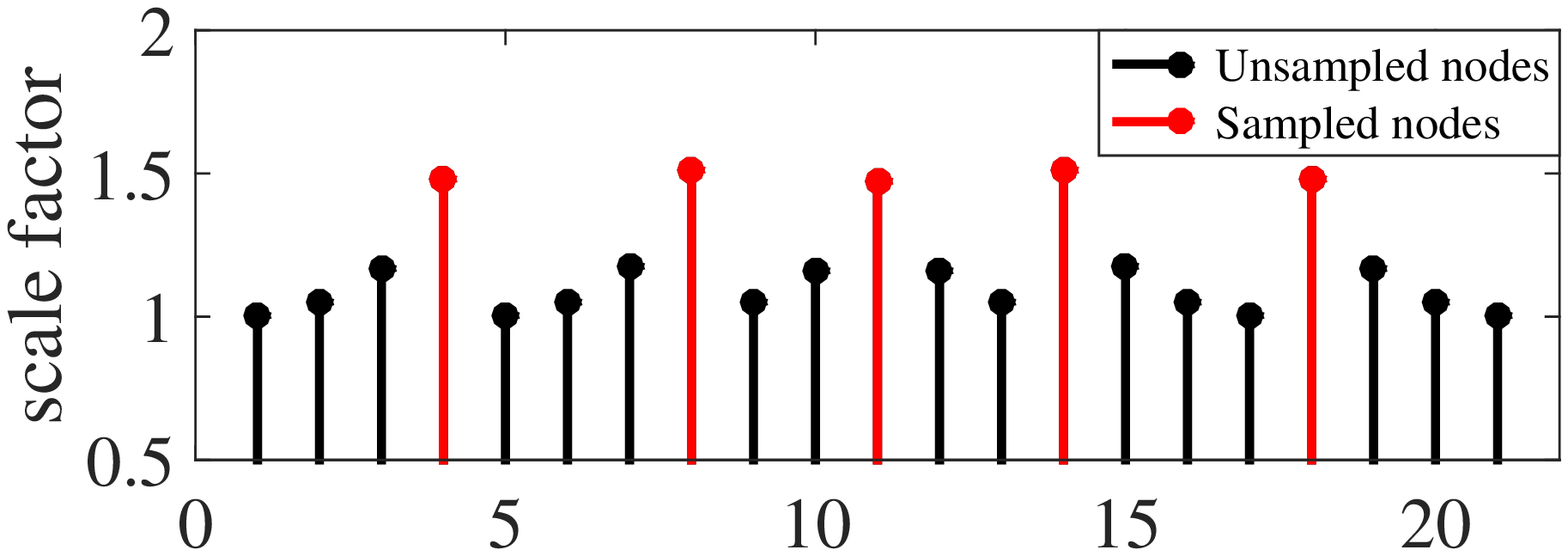}}
\subfloat[]{
\label{fig:lg_bf7}
\includegraphics[width=0.49\linewidth]{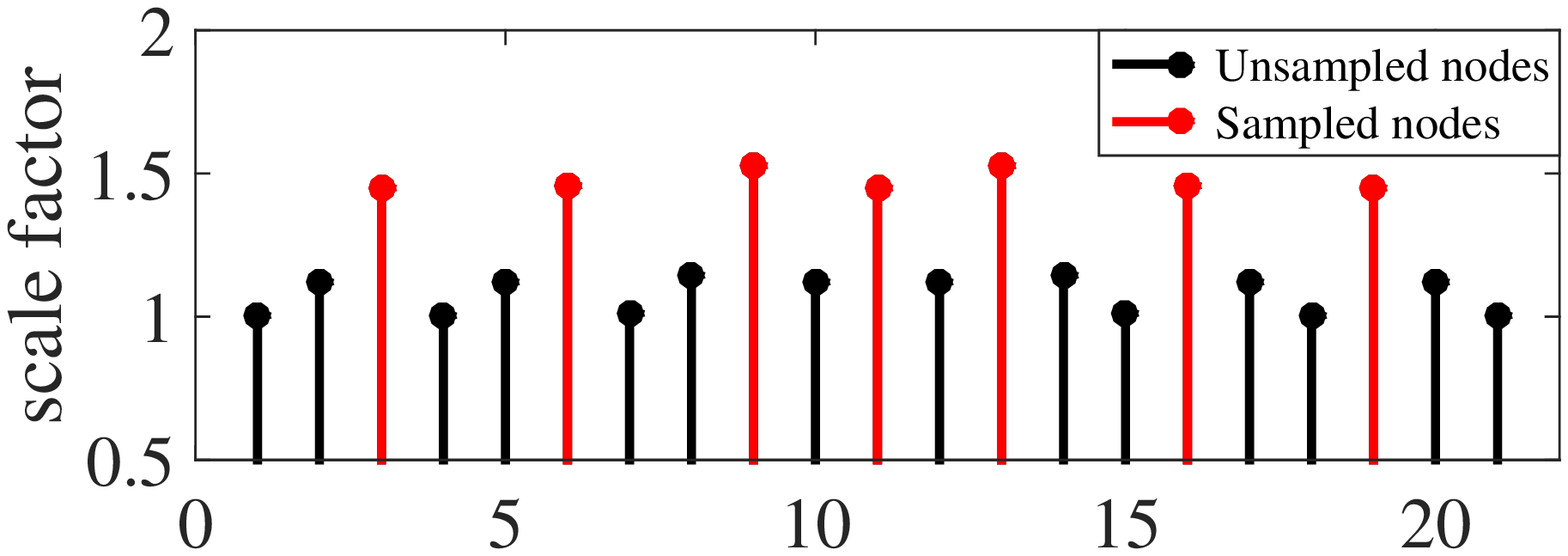}}
\caption{Sampling on an unweighted line graph, $|\cV|=21$. (a) Sampling $5$ nodes, lower-bound $T=0.048$. (b) Sampling $7$ nodes, lower-bound $T=0.107$.}
\label{fig:line_graph}
\end{figure}

\begin{figure}[!t]
\centering
\subfloat[]{
\label{fig:lb}
\includegraphics[width=0.49\linewidth]{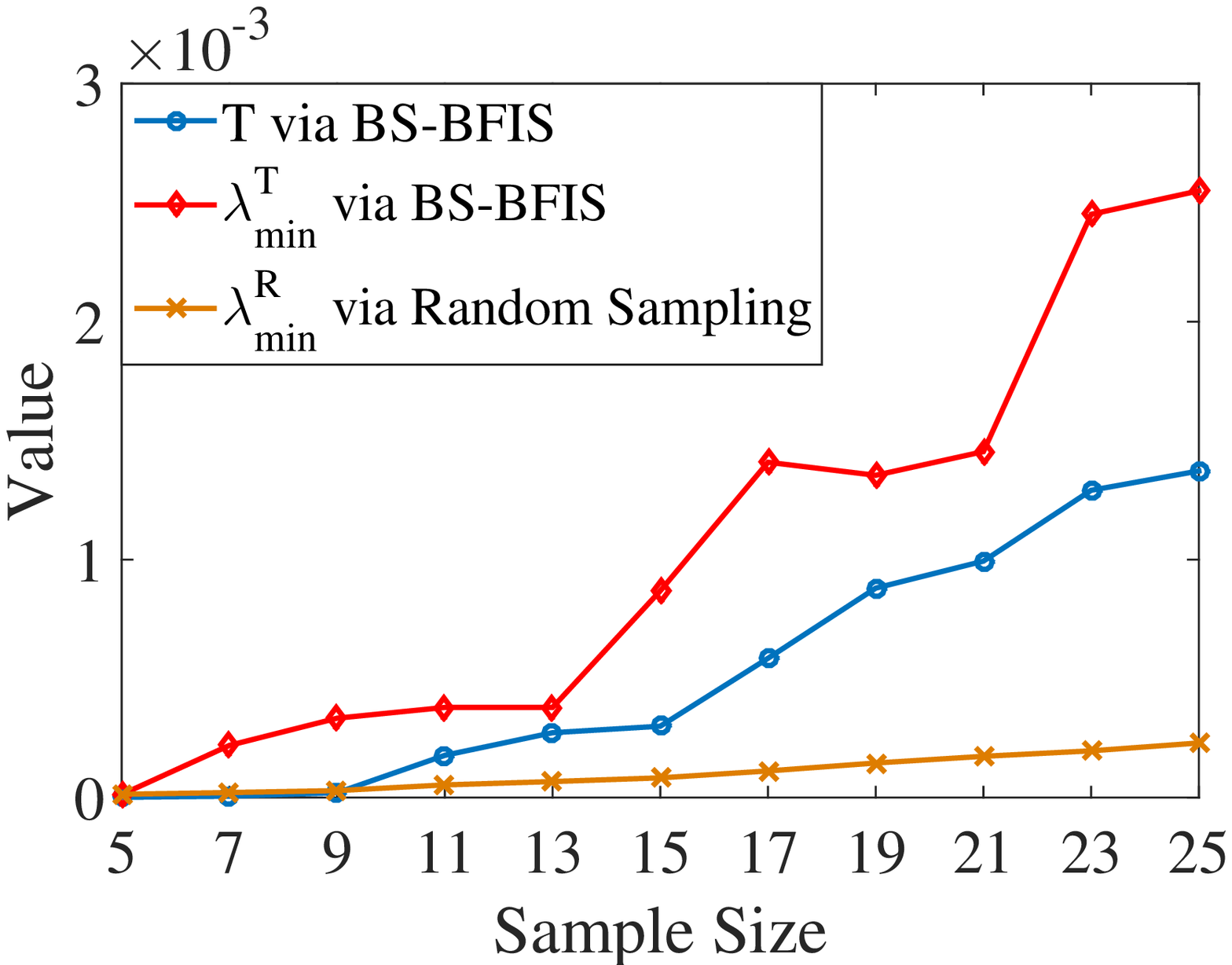}}
\subfloat[]{
\label{fig:mse}
\includegraphics[width=0.49\linewidth]{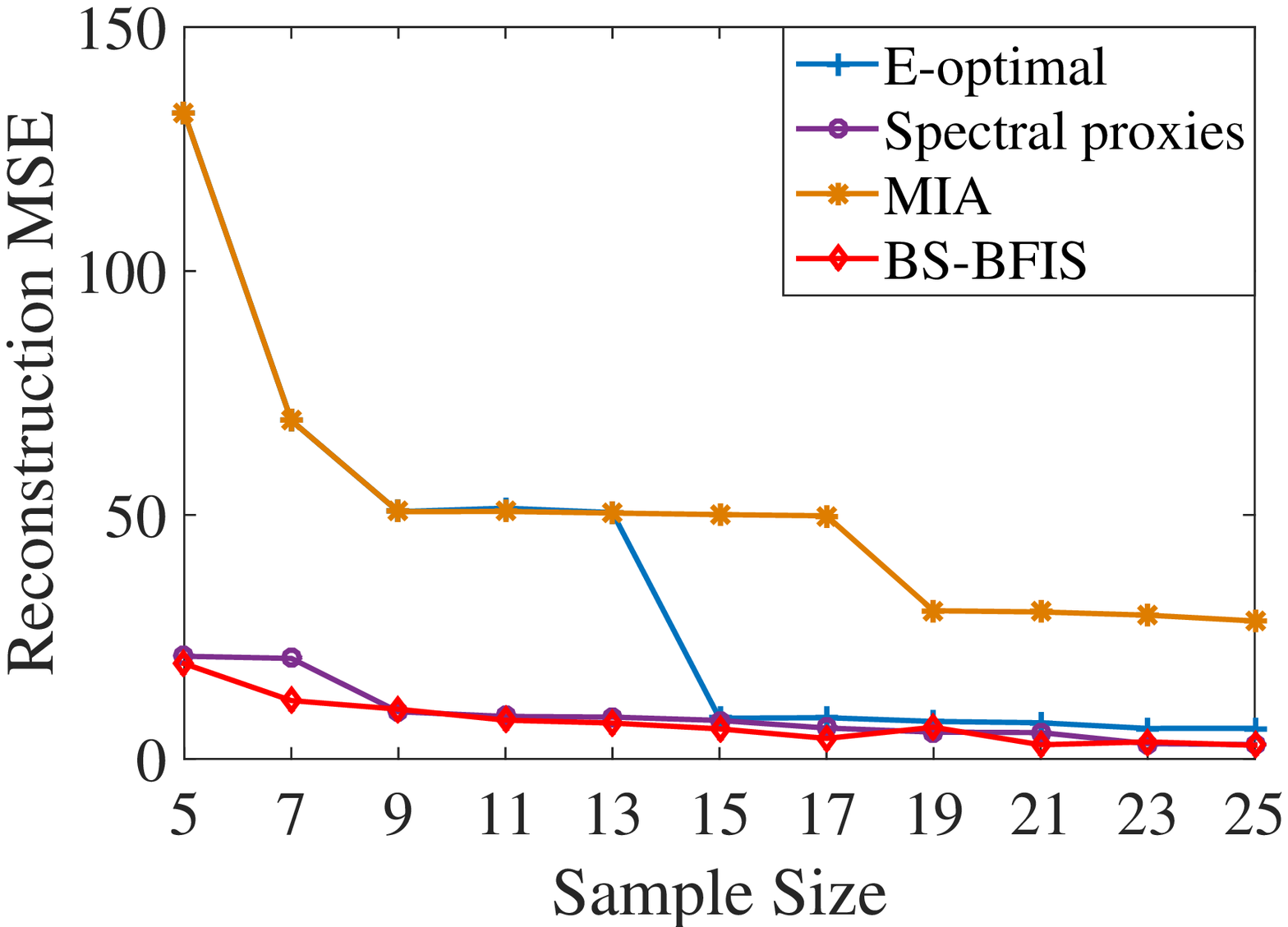}}
\caption{(a) Comparisons among lower-bound $T$ and the corresponding $\lambda^T_{\min}$ via BS-BFIS, and mean $\lambda^R_{\min}$ via 100 times random sampling. (b) Reconstruction MSE comparisons among E-optimal \cite{e_optimal2015}, Spectral proxies \cite{sp_proxy2016}, MIA \cite{MIA2018Fen} and BS-BFIS.}
\label{fig:climate_quantitative}
\end{figure}

For experiments on real data, we build a graph on real \textit{U.S. Climate Normals database} \cite{climate_database}. We select $100$ temperature stations close to cities with $100$ largest populations as graph nodes.
The graph edges are connected with \textit{Delaunay Triangulation}\footnote{https://en.wikipedia.org/wiki/Delaunay\_triangulation}, and the graph weights are computed using $w_{ij}=\exp(-\|\l_i-\l_j\|^2_2/\sigma_l^2)\cdot\exp(-\|\x_i-\x_j\|^2_2/\sigma_x^2)$ like bilateral filter \cite{tomasi98}, where $\l_i$ and $\x_i$ are the geometric location and the temperature of station $i$, respectively. $\sigma_l=5$ and $\sigma_x=3$. In our experiments, we sample the temperatures of $K$ stations with simulated additive Gaussian noise of unit variance. Then, we reconstruct temperatures of all stations by solving linear equation (\ref{eq:linear_equation}).

\subsection{Experimental Results}
In Fig.\;\ref{fig:line_graph}, we conduct an illustrative experiment to perform sampling on an unweighted line graph of $21$ nodes.
We sample $5$ and $7$ nodes, respectively. Fig.~\ref{fig:lg_bf} and Fig.~\ref{fig:lg_bf7} report the scale factor $s_i$ for each disc and the distribution of sampled nodes.
Using BS-BFIS, we observe periodic uniform sampling for different sampling budgets, which agrees with our intuition.

We also apply BS-BFIS on a graph built on real \textit{U.S. Climate Normals database} \cite{climate_database}.
Our objective is to maximize the lower-bound of minimum eigenvalue $\lambda_{\min}$.
We apply BS-BFIS on the constructed graph $\cG$ to compute the lower-bound threshold $\hT$ and sampling matrix $\A$ with increasing sample budget $K$.
With output $\A$, we compute $\lambda^T_{\min}$ via eigen-decomposition.
For comparison, we employ random sampling 100 times and compute the mean minimum eigenvalue $\lambda^R_{\min}$.
As shown in Fig.\;\ref{fig:lb}, BS-BFIS can promote large lower-bound threshold $T$ with increasing sample budget $K$, and the minimum eigenvalue $\lambda^T_{\min}$ increases correspondingly. Both the lower-bound $T$ and the corresponding $\lambda^T_{\min}$ increases much faster than $\lambda^R_{\min}$ using random sampling.

\begin{figure}[!t]
\centering
\subfloat[E-optimal \cite{e_optimal2015}.]{
\label{fig:c_e}
\includegraphics[width=0.47\linewidth]{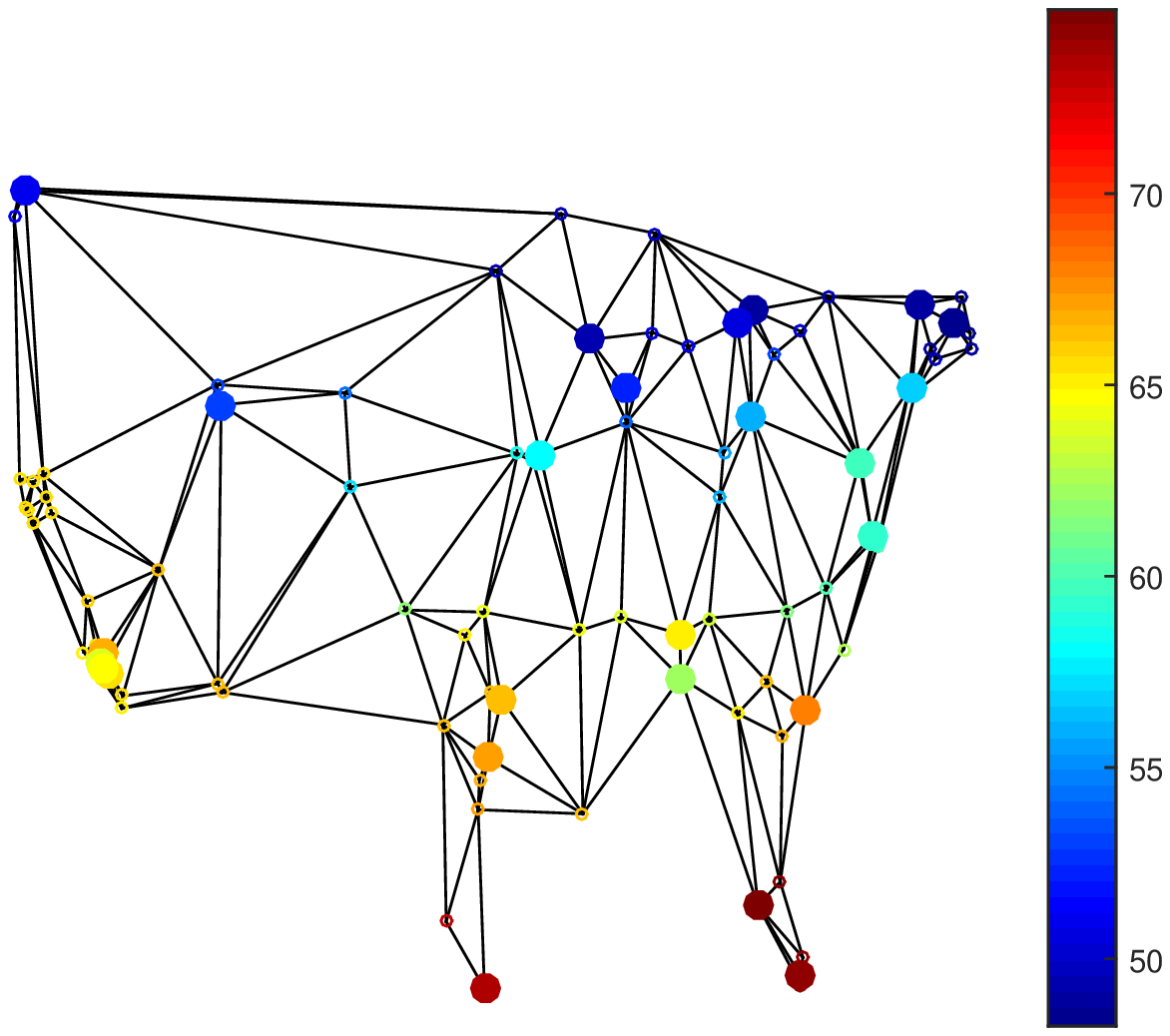}}
\subfloat[Spectral proxies \cite{sp_proxy2016}.]{
\label{fig:c_ao}
\includegraphics[width=0.47\linewidth]{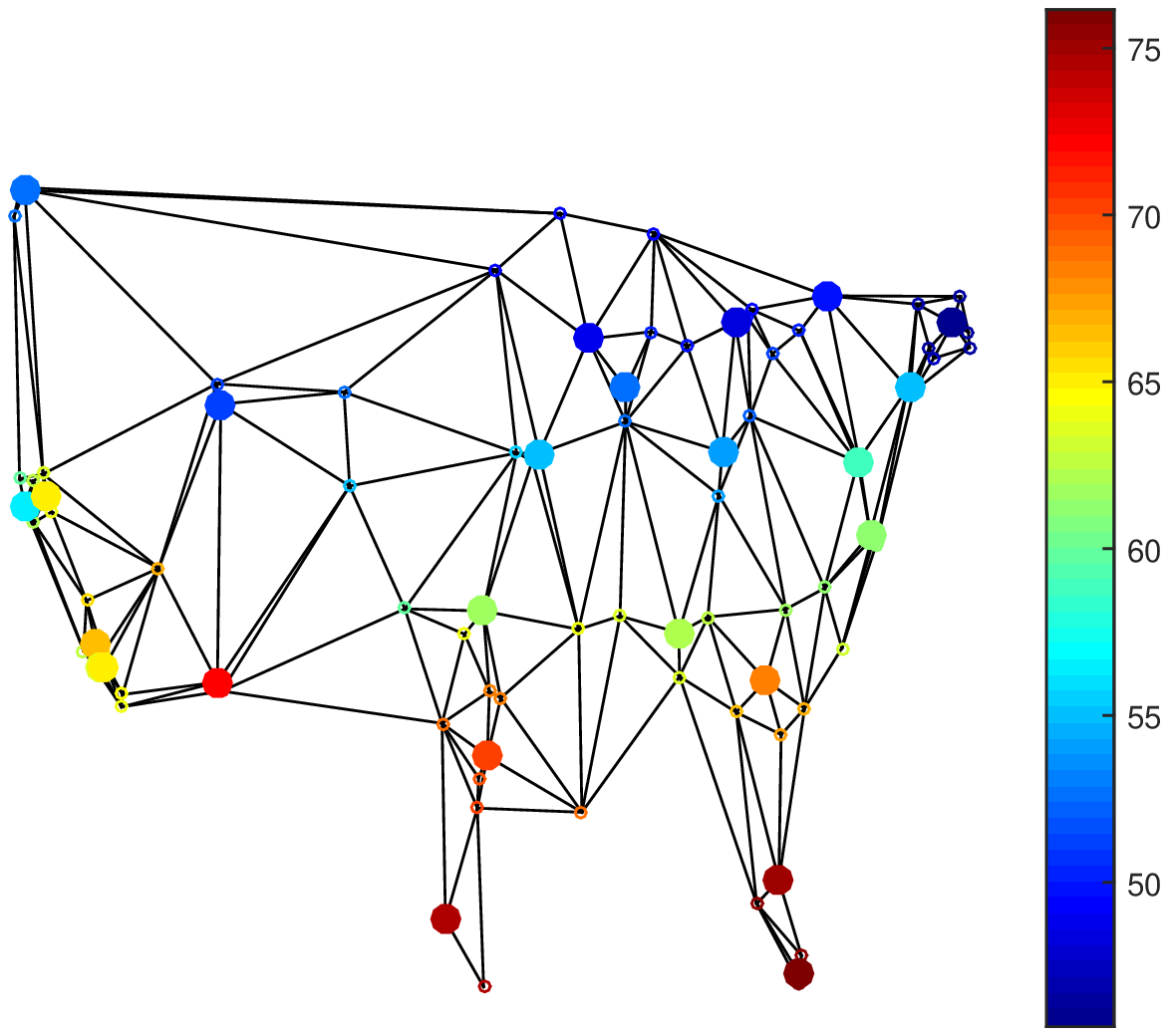}} \\
\subfloat[MIA \cite{MIA2018Fen}.]{
\label{fig:c_a}
\includegraphics[width=0.47\linewidth]{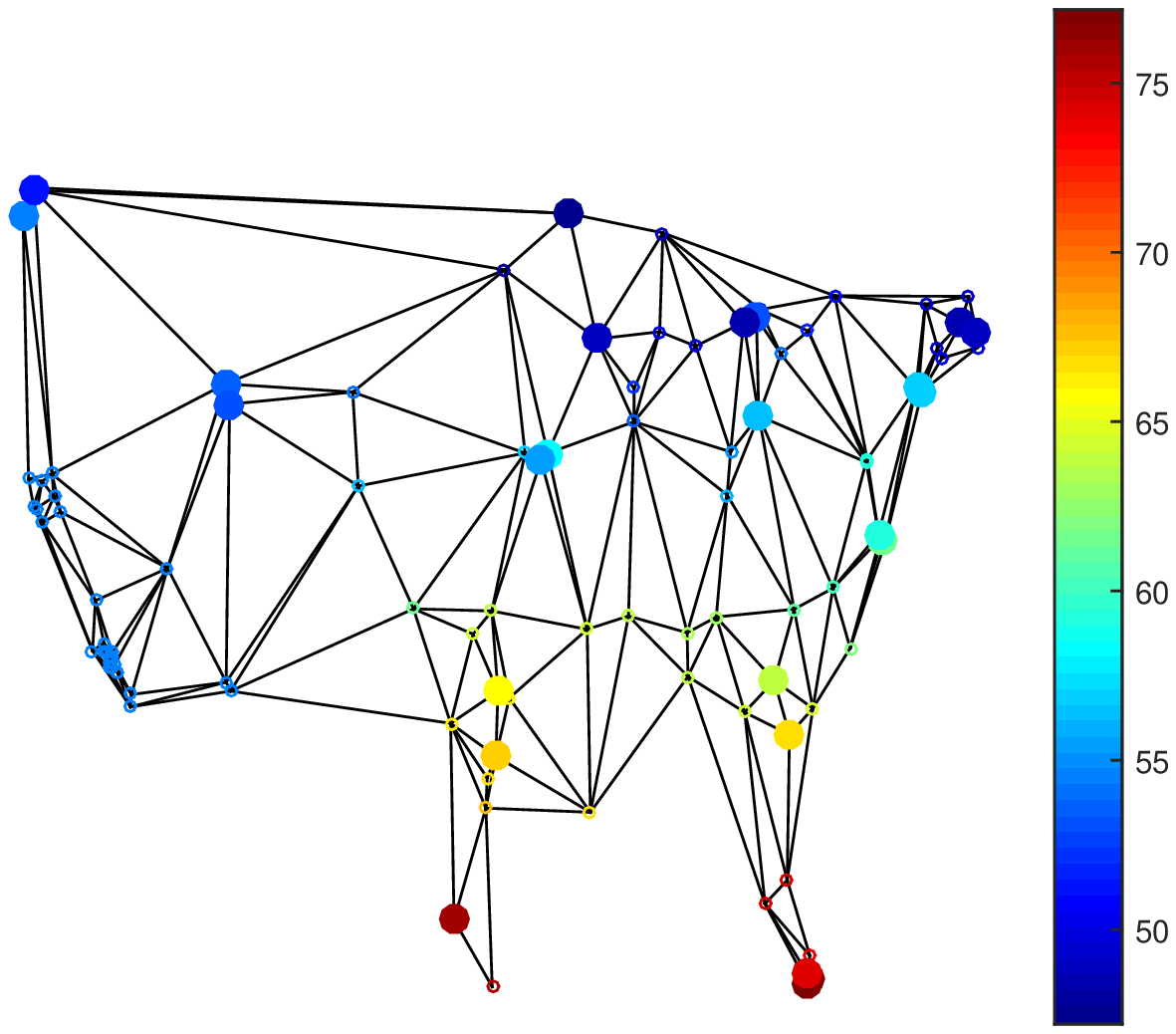}}
\subfloat[BS-BFIS.]{
\label{fig:c_bs}
\includegraphics[width=0.47\linewidth]{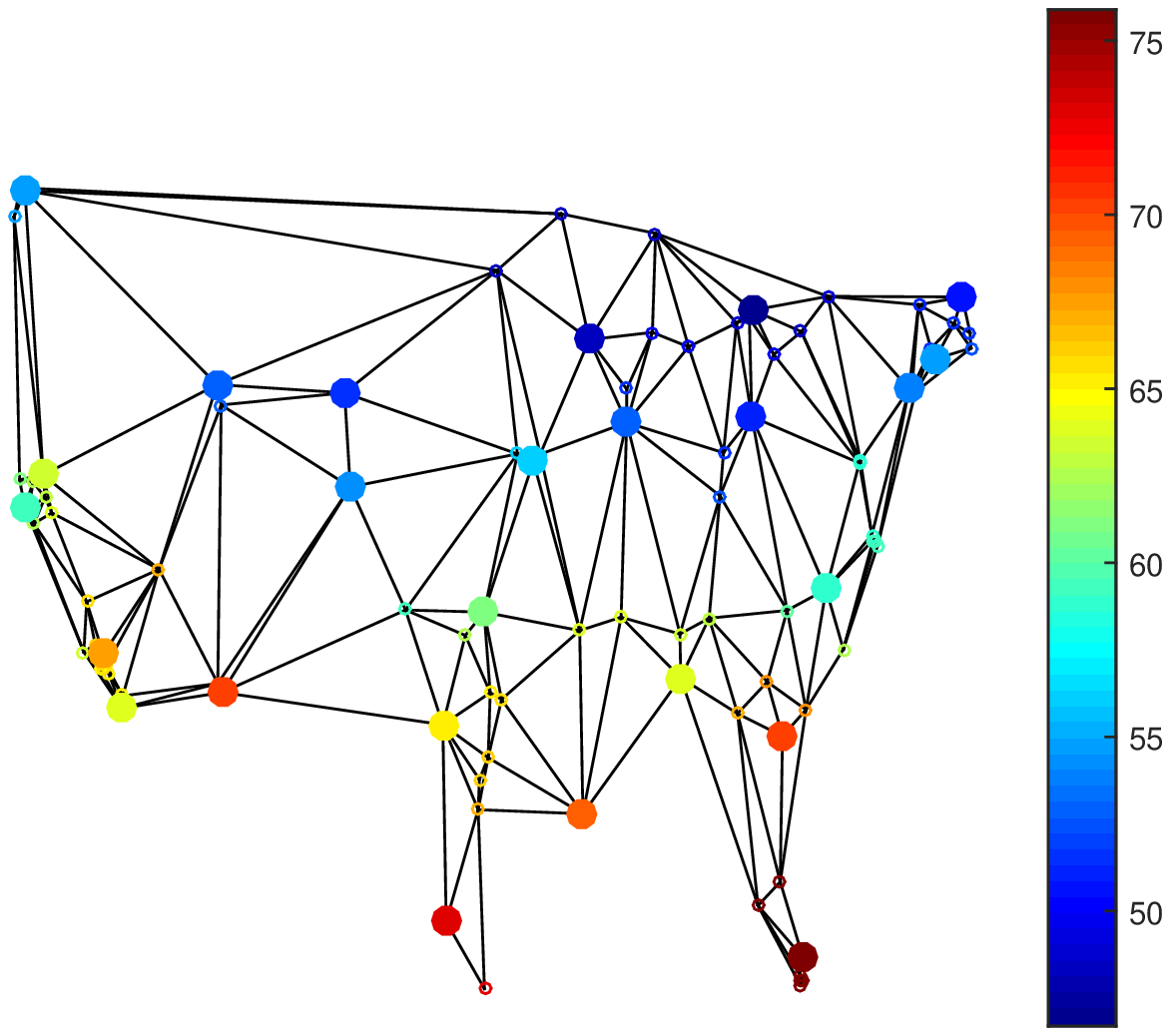}}
\caption{Sampling visualization ($K=25$). Solid circles are sampled nodes. Color depicts the temperature. The running time of E-optimal \cite{e_optimal2015}, Spectral proxy \cite{sp_proxy2016}, MIA \cite{MIA2018Fen} and BS-BFIS is $0.103s$, $1.440s$, $0.108s$ and $0.082s$, respectively.}
\label{fig:climate}
\end{figure}

We also compare the reconstruction MSE of BS-BFIS with existing sampling methods: E-optimal \cite{e_optimal2015}, spectral proxies \cite{sp_proxy2016}, and MIA \cite{MIA2018Fen}, as shown in Fig.~\ref{fig:mse}.
Each method outputs sampling matrix $\A$ under sampling size $K$. With $\A$, we can have $\H$ and solve (\ref{eq:linear_equation}) to reconstruct the temperatures of all stations.
We observe that the performance of BS-BFIS is comparable to or better than the competing methods.
In Fig.\;\ref{fig:climate}, we visualize the sampled nodes of the four methods with $K=25$ and show the running time, respectively.
We observe that the sampled nodes of BS-BFIS tend to distribute uniformly on the graph, due to BFS and disc scaling operation in BFIS.
However, sampled nodes of other methods, such as MIA \cite{MIA2018Fen}, tend to accumulate in several areas.
This explains the good performance of BS-BFIS. BS-BFIS is the fastest among the four algorithms.

\vspace{-0.1in}